\documentclass[
  reprint,
  amsmath,amssymb,
  aps
]{revtex4-2}

\usepackage[english]{babel}
\usepackage{ulem}
\makeatletter
\@namedef{l@en}{\l@english}
\makeatother

\usepackage[dvipsnames]{xcolor}
\usepackage{graphicx}
\usepackage{dcolumn}
\usepackage{bm}
 \usepackage{multirow}
\usepackage[table,xcdraw]{xcolor}

\begin{document}

\title{\bf Inferring the microscopic mechanisms of opinion dynamics using a kinetic Ising model}
\thanks{ixandra.achitouv@cnrs.fr}%

\author{Ixandra Achitouv}
 \affiliation{Sorbonne Université, CNRS, LIP6, F-75005 Paris, France}
\affiliation{Complex Systems Institute of Paris île-de-France (ISC-PIF, UAR3611), Paris, France}

\author{David Chavalarias}%

\affiliation{Complex Systems Institute of Paris île-de-France (ISC-PIF, UAR3611), Paris, France}%
 
\affiliation{Centre d'Analyse et de Mathématique Sociales (CAMS, UMR8557), Paris, France}%

\author{Vincent Lahoche}
 \affiliation{Université Paris-Saclay, CEA, Palaiseau, F-91120, France}


\begin{abstract}
Kinetic Ising models are widely used to describe binary opinion dynamics, but their microscopic validity has rarely been tested empirically. Here, we infer the transition probabilities governing opinion updates from a year-long online social network and show that they are accurately described by an Ising heat-bath dynamics. The inferred parameters admit a direct sociological interpretation: the external field quantifies intrinsic bias, the coupling strength measures social influence, and a persistence term captures temporal inertia. We further show that persistence is positively correlated with node degree, while both persistence and interaction strength are strongly correlated with global network heterogeneity and clustering. Using the inferred time-dependent parameters in Monte Carlo simulations on the empirical temporal networks, we accurately reproduce the observed response functions, flip probabilities, and macroscopic opinion dynamics. Within the Twitter climate debate, our results provide direct empirical support for a kinetic Ising description of online opinion formation and establish a quantitative link between microscopic social behavior and evolving network topology.
\end{abstract}

\maketitle

\section{Introduction}

Kinetic Ising models have long provided a powerful statistical mechanics framework for understanding collective behavior, phase transitions, and emergent phenomena \cite{ising_beitrag_1925,glauber1963}. Beyond physics, they have become a cornerstone of computational social science for modeling binary opinion dynamics, where individuals update their opinions under the combined effects of social influence and intrinsic biases \cite{castellano2009,peralta2022opiniondynamicssocialnetworks}. In these models, the macroscopic polarization of a population is naturally interpreted as a magnetization process emerging from the microscopic transition probabilities of interacting agents.

Numerous opinion-dynamics models have been proposed over the last decades, including consensus models \cite{degroot_reaching_1974,friedkin_social_1990}, voter-type dynamics \cite{holley_ergodic_1975}, bounded-confidence models \cite{deffuant_mixing_2000,hegselmann2002}, Galam models \cite{galam_contrarian_2004}, and kinetic Ising approaches \cite{castellano2009}. However, despite their widespread theoretical use, the microscopic validity of kinetic Ising models has rarely been tested against high-resolution temporal observations of real online social systems. Most studies postulate interaction rules or transition kernels a priori, leaving a significant gap between idealized theoretical models and the actual mechanisms governing opinion changes in digital communication environments \cite{castellano2009,peralta2022opiniondynamicssocialnetworks}.

In this work, we bridge this gap by inferring the microscopic transition probabilities governing opinion updates directly from a year-long dataset of a large-scale online social network. By tracking the evolution of the climate-change debate on Twitter throughout 2022, we analyze the fine-grained temporal dynamics of opinion shifts. We demonstrate that the empirical probability of opinion changes is remarkably well described by a kinetic Ising heat-bath (Glauber-like) dynamics.

Crucially, our data-driven approach allows us to disentangle distinct social mechanisms by testing a hierarchy of kinetic Ising models. We show that the inferred transition parameters admit direct sociological interpretations: the external field captures intrinsic biases, the interaction strength quantifies peer influence, and a persistence term accounts for temporal inertia. By integrating these inferred, time-dependent parameters into Monte Carlo simulations, we successfully reproduce the empirical response functions and the macroscopic opinion trajectories observed in the data.

Furthermore, previous studies have demonstrated that network topology strongly influences collective (macroscopic) opinion dynamics, including consensus formation, polarization, and the emergence of extremism
\cite{amblard_role_2004,Sasahara_2020,Liu_2023,perrier_phase_2024}.
Here, we establish a complementary microscopic result. 
Rather than simply affecting the collective outcome of the dynamics, we observe a strong
interdependence between network topology and the microscopic transition kernel,
whose inferred parameters capture individual behavioral traits within the
population. In particular, opinion persistence is not homogeneous across
individuals but is strongly associated with network connectivity, with highly
connected users exhibiting substantially greater resistance to opinion changes.
Our analysis does not establish the direction of causality: persistent
individuals may generate specific network structures, while these structures
may in turn influence subsequent opinion dynamics.

 More broadly, this work provides an empirical validation of the kinetic Ising description of online opinion formation. It establishes a quantitative link between the structural properties of evolving communication networks and the microscopic kernels of social influence, providing a physically interpretable framework for studying the emergence, persistence, and regulation of collective opinions in digital social systems.

The remainder of this article is organized as follows. Section~\ref{sec:data} describes the Twitter dataset, the preprocessing pipeline, and the temporal evolution of the interaction network. Section~\ref{sec:Mod} introduces the hierarchy of kinetic Ising models together with the parameter inference procedure. Section~\ref{sec:MCMC} presents Monte Carlo simulations of the inferred kinetic Ising models, compares their predictions with the empirical observations and discusses the inferred microscopic mechanisms and their relationship with the evolving network topology. Finally, Section~\ref{sec:conclu} summarizes the main findings and discusses their implications for the modeling of online opinion dynamics.

\section{Datasets Analysis}\label{sec:data}

\subsection{Methodology}

The empirical data are drawn from the \textit{Climatoscope} project~\cite{chavalarias_hashtags_2025}, which continuously collected Twitter messages related to climate change through the Twitter Track API using several dozen English and French climate-related keywords. The resulting dataset comprises approximately 57 million tweets posted during 2022, 32.1 million of those are retweets, and captures the large-scale online debate between pro-climate and climate-denialist communities.

{Recurring retweets are good indicators of opinion alignment \cite{gaumont_reconstruction_2018}. In this paper, we study the contribution of the retweet graph to the opinion dynamics following the methodology introduced in ~\cite{achitouv2026dmodddiffusionmodelopinion}. We first reconstruct a sequence of weighted retweet} networks using a sliding time window of three weeks shifted every seven days, yielding 50 temporal interaction networks over the year.  Nodes represent Twitter accounts, while weighted edges correspond to the number of retweets exchanged between users. To remove spurious interactions, only edges with a weight larger than the median retweet count ($>1$) within each time window are retained, resulting in temporal networks involving approximately $6.9\times10^5$ unique users. {Last, we filter the users that are present at least 5 periods to end with a temporal network involving 165.4k accounts.} 

To infer users' opinions, each temporal network is embedded using the \texttt{node2vec} algorithm~\cite{grover2016node2vec}, followed by a supervised two-dimensional UMAP projection~\cite{mcinnes2018umap} trained on two sets of reference accounts with well-established pro-climate and climate-denialist positions. The resulting one-dimensional opinion coordinate \(x_i(t)\), introduced in ~\cite{achitouv2026dmodddiffusionmodelopinion}, is subsequently coarse-grained into Ising spin variables,
\[
s_i(t)\in\{-1,+1\},
\]
according to the sign of the inferred opinion,
\[
s_i(t)=
\begin{cases}
+1,& x_i(t)>0,\\
-1,& x_i(t)<0.
\end{cases}
\]
Here, \(x_i=0\) corresponds to the decision boundary separating the pro-climate and climate-denialist communities. Although the latent opinion variable is continuous, its empirical distribution is strongly bimodal, with the vast majority of users concentrated around two well-separated attractors located near \(x\simeq\pm2\) (see Ref.~\cite{achitouv2026dmodddiffusionmodelopinion}). Consequently, only a small fraction of users lie close to the decision boundary. This strong polarization justifies the use of the binary Ising representation as a natural effective two-state description of the underlying opinion space. Throughout this work, we therefore model the temporal evolution of opinions as a stochastic process acting on these binary spin variables.

\subsection{Dynamical properties of the interaction network}\label{sec:Network}
\begin{figure}
    \centering
    \includegraphics[width=1\linewidth]{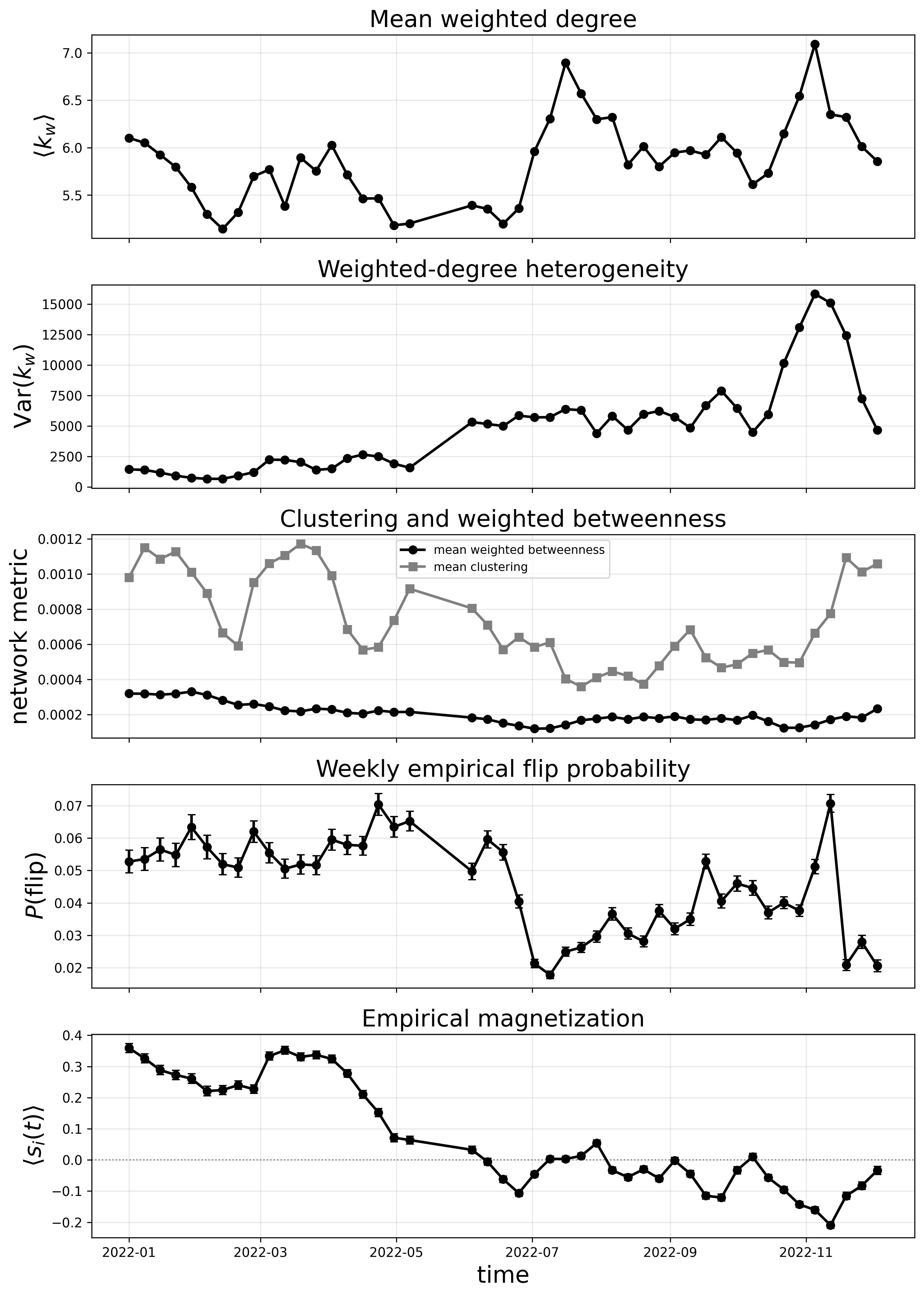}
\caption{Temporal evolution of the interaction network during the 2022 climate-change debate, the empirical weekly probability of opinion changes $P(\mathrm{flip})$, and the average opinion (magnetization) $\langle s_i(t)\rangle$.}
    \label{fig:network_dynamics}
\end{figure}

Before inferring microscopic opinion dynamics, we first characterize the temporal evolution of the underlying interaction network. 

{On the same dataset, \cite{chavalarias_hashtags_2025} have shown that there was a sharp surge in climate denial on Twitter in July 2022, both in terms of the number of active accounts and the number of tweets per account, alongside a strengthening of this community’s structure and the emergence of super-spreaders. This had been preceded a few months earlier by an increase in the proportion of likely automated accounts in the denialist community.}

{Here we further quantify this evolution by characterizing the micro-dynamics of opinion evolution which gives insights into the nature of these new accounts. Since we use the retweet network as a proxy for the local social field experienced by each individual}, changes in the network topology are expected to {be correlated with} the collective opinion dynamics. Figure~\ref{fig:network_dynamics} summarizes the evolution of several global structural observables over the course of 2022.

The upper panel shows the evolution of the mean weighted degree,
\[
\langle k_w\rangle=\frac{1}{N}\sum_i k_i,
\]
which measures the average interaction strength of users. The network remains relatively stable during the first half of the year, with an average weighted degree between 7 and 8. A marked increase is observed during the summer 2022, where the average weighted degree reaches values above 10, before returning to intermediate values toward the end of the year. 

The second panel reports the variance of the weighted degree distribution,
\[
\mathrm{Var}(k_w),
\]
which quantifies the heterogeneity of the interaction network. During the first months of the year the network is relatively homogeneous, whereas from June onward the variance increases, exceeding $10^4$ during the autumn. This increase reveals the emergence of highly connected hub accounts that concentrate a growing fraction of the information flow.

The third panel presents two complementary measures of global network organization: the average weighted betweenness centrality and the average clustering coefficient. Weighted betweenness decreases during the first half of the year and remains comparatively low throughout the summer, indicating that shortest communication paths become distributed over a larger set of users. In contrast, the clustering coefficient exhibits a non-monotonic evolution, with higher values during the beginning and end of the year and a pronounced minimum during the summer months, {which is in line with the structuration of the denialist community around few 'star' accounts and its relative growth compared to the pro-climate community}.

The fourth panel reports the empirical weekly flip probability, 
\begin{equation}
P(\mathrm{flip})=\frac{1}{N_t}\sum_{i=1}^{N_t}
\mathbf{1}\left[s_i(t+1)\neq s_i(t)\right],
\end{equation}
computed as the fraction of users whose binary opinion changes between two consecutive weekly observations. The flip probability decreases markedly during the second half of the year, coinciding with the increase in the average weighted degree and the weighted-degree heterogeneity, and with the reduction of both clustering and weighted betweenness.

\subsection{{Temporal networks and turnover in climate activists}\label{sec:turnover}}
The above observations reveal roughly two distinct dynamical regimes for climate change opinion separated by a transition period. From January to March 2022 (period 1, 56.8k accounts), there is a majority of pro-climate accounts (empirical average magnetization $> 0.3$), a low heterogeneity in weighted degree, higher clustering and high empirical flip probability ($\approx2,5\%$ between periods). From July to Septembrer 2022 (period 2, 114k accounts), there is an even population of pro-climate and skeptics, a high weighted degree heterogeneity, lower clustering and a low empirical flip probability ($\approx1,3\%$ between periods).

The analysis of accounts turnover reveals that the second period is composed of 36,6k persistent accounts from period 1 (including 13,8k who are denialists) and 77,4k new accounts (including 46,2k who are denialists). There is consequently a high turnover between the two periods and a large influx of denialists accounts. 

The mean degree of persistent pro-climate accounts has increased from 7.66 to 9.25 ($+20\%$) between the two periods while the one of denialists has increased from 5.15 to 12.3 ($+139\%$). On the other hand, the mean degree of new pro-climate accounts is 3.15 vs. 2.94 for denialists. This means that denialist newcomers are entering the arena to support existing accounts rather than to contribute to a debate. 

It is particularly clear when analysing the flip probabilities. For persistent accounts,  for both periods, denialist accounts have a higher probability to become pro-climate than pro-climate have to become denialists, and the flip probability of both population decrease on average with time, suggesting that their members are on average more confident in their opinions with time. But for period 2, denialists newcomers even less likely to change their mind than any other population and about less than twice than pro-climate newcomers (see table~\label{table:flipStats}). This confirms that denialist newcomers are not there to make an opinion but to defend their convictions.

\begin{table}[]
\caption{Evolution of individual characteristics between the two periods P1 and P2.\label{table:flipStats}}
\label{tab:my-table}
\begin{tabular}{|clllll|}
\hline
\multicolumn{1}{|l}{\textbf{}}                                                       & \multicolumn{1}{l|}{\textbf{}}                           & \multicolumn{2}{c|}{\textbf{Proclimate}}          & \multicolumn{2}{c|}{\textbf{Denialists}}                   \\ \hline
\rowcolor[HTML]{ECF4FF} 
\multicolumn{1}{|c|}{\cellcolor[HTML]{ECF4FF}}                                       & \multicolumn{1}{l|}{\cellcolor[HTML]{ECF4FF}\textbf{P1}} & \multicolumn{2}{c|}{\cellcolor[HTML]{ECF4FF}7.66} & \multicolumn{2}{c|}{\cellcolor[HTML]{ECF4FF}5.15}          \\
\rowcolor[HTML]{ECF4FF} 
\multicolumn{1}{|c|}{\multirow{-2}{*}{\cellcolor[HTML]{ECF4FF}\textbf{Mean degree}}} & \multicolumn{1}{l|}{\cellcolor[HTML]{ECF4FF}\textbf{P2}} & \multicolumn{2}{c|}{\cellcolor[HTML]{ECF4FF}9.25} & \multicolumn{2}{c|}{\cellcolor[HTML]{ECF4FF}\textbf{12.3}} \\ \hline
\multicolumn{1}{|c|}{}                                                               & \multicolumn{1}{l|}{\cellcolor[HTML]{FFFFFF}\textbf{P1}} & 4.13   & \multicolumn{1}{l|}{{[}4.1 ; 4.15{]}}    & 6.57                        & {[}6.52 ; 6.63{]}            \\
\multicolumn{1}{|c|}{\multirow{-2}{*}{\textbf{p(flip) {[}persistent{]}}}}            & \multicolumn{1}{l|}{\cellcolor[HTML]{FFFFFF}\textbf{P2}} & 2.89   & \multicolumn{1}{l|}{{[}2.88 ; 2.91{]}}   & 3.5                         & {[}3.48 ; 3.52{]}            \\ \hline
\rowcolor[HTML]{ECF4FF} 
\textbf{p(flip) {[}newcomers{]}}                                                     & \textbf{P2}                                              & 4.38   & {[}4.35 ; 4.4{]}                         & \textbf{2.22}               & {[}2.21 ; 2.23{]}            \\ \hline
\end{tabular}
\end{table}

The populations of period 1 and period 2 are consequently of very different nature, with a wave of stubborn denialists accounts that entered the game to amplify persistent denialists accounts, whose local network became star-like. Such a mechanism naturally increases degree heterogeneity while decreasing the clustering coefficient.  

Could these the populations of period 1 and period 2 be described by the same opinion dynamics model despite these striking differences? In that case,  what can we learn from the parameters of the model?


Within the kinetic Ising framework, the flip probability provides a direct measure of the effective level of stochasticity of the dynamics. In a heat-bath dynamics, large flip probabilities correspond to a high-temperature regime in which thermal fluctuations dominate the interaction energy, whereas small flip probabilities indicate a low-temperature regime where spins are more likely to remain aligned with their local effective field. The observed reduction of $P(\mathrm{flip})$ therefore suggests that the climate discussion progressively evolves toward a more ordered dynamical regime, characterized by stronger effective social interactions and increased temporal stability of individual opinions. This interpretation is consistent with the subsequent inference of the kinetic Ising parameters discussed in sec.\ref{sec:Mod}.

Interestingly, the bottom panel reports the average opinion (magnetization),
\[
\langle s_i(t)\rangle=\frac{1}{N_t}\sum_{i=1}^{N_t}s_i(t),
\]
which measures the global polarization of the discussion. Throughout period 1, the magnetization remains predominantly positive, indicating the dominance of the pro-climate opinion in the sampled population, while exhibiting noticeable temporal fluctuations associated with major events, while it falls negative around 0 in period 2.

{In terms of the Ising formalism, we would say that the denialist newcomers contribute to create a more ordered state in which individual opinions become increasingly stable.}

Furthermore, in the kinetic Ising framework, this behavior is characteristic of a regime with stronger effective social interactions (equivalently, a lower effective temperature), where alignment with the local social environment dominates over stochastic opinion changes.

Overall, the climate discussion network undergoes substantial structural changes throughout 2022. In particular, the simultaneous increase in the average weighted degree and its variance indicates periods of intensified activity accompanied by a progressively more heterogeneous interaction structure. 

\subsection{Network properties \& opinion dynamics}

\subsubsection*{Local interactions \& persistence}
\begin{figure}
    \centering
    \includegraphics[width=1\linewidth]{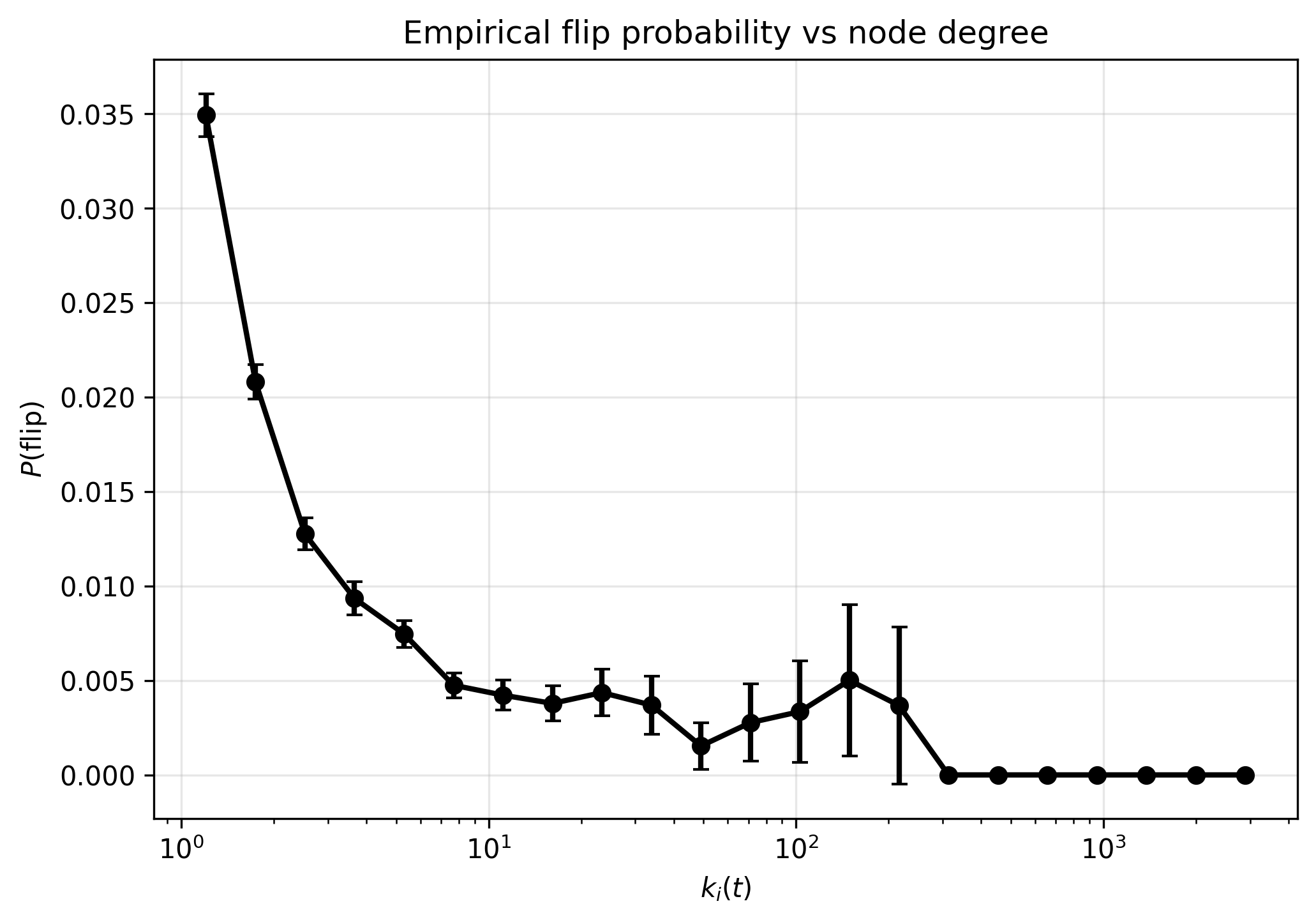}
\caption{
Empirical probability of opinion change as a function of the weighted node degree $k_i(t)$. Nodes are grouped into logarithmically spaced degree bins, and the markers denote the mean flip probability within each bin. Error bars indicate $95\%$ binomial confidence intervals.}
    \label{fig:flip_vs_degree}
\end{figure}

 Figure~\ref{fig:flip_vs_degree} reports the empirical probability that a user changes opinion between two consecutive weeks as a function of its weighted degree.

A clear monotonic trend emerges. Users with very small weighted degree exhibit a relatively large probability of changing opinion, exceeding $3\%$ for isolated nodes. The flip probability then decreases rapidly with increasing degree and falls below $0.5\%$ for moderately connected users. For the most connected accounts, opinion changes become extremely rare, with virtually no observed flips despite the large number of interactions.

When aggregated over the full year, opinion persistence is strongly associated
with connectivity, with highly connected users changing opinion less frequently
than peripheral users. This empirical observation motivates the introduction of Model M4 we introduce in sec.\ref{sec:Mod}. This relationship, however, should not be interpreted as evidence that network position causally determines
individual persistence. In particular, the influx of highly connected denialist
users during the second part of the year modifies the composition of the
population and the relation between connectivity and persistence. Rather than
describing individuals as agents with fixed properties, the kinetic Ising
framework provides a common functional description across these different
regimes, while changes in its inferred parameters summarize the evolving
balance between individual traits, connectivity, and opinion persistence.

\subsubsection*{Network topology {\& opinion dynamics}}
\begin{figure}
    \centering
    \includegraphics[width=1\linewidth]{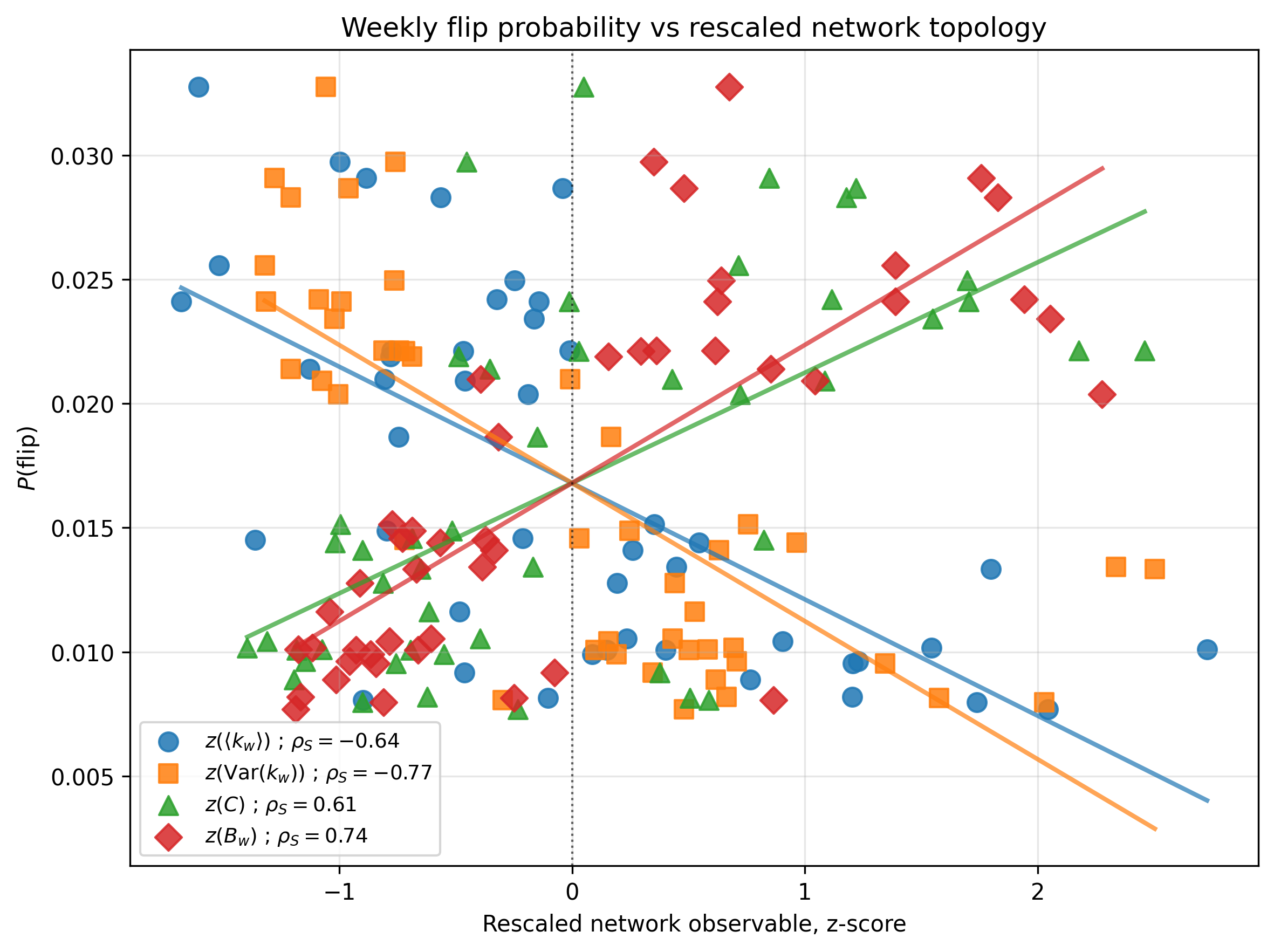}
    \caption{Relationship between the empirical weekly opinion flip probability and standardized global network observables. Each point corresponds to one weekly retweet network. The horizontal axis shows the $z$-score of four network diagnostics: mean weighted degree $\langle k_w\rangle$ (blue circles), weighted-degree heterogeneity $\mathrm{Var}(k_w)$ (orange squares), mean clustering coefficient $C$ (green triangles), and mean weighted betweenness centrality $B_w$ (red diamonds). Solid lines denote linear regressions for visual guidance, and the legend reports the corresponding Spearman correlation coefficients.}
    \label{fig:Pflip_netz}
\end{figure}

The microscopic analysis above demonstrates that highly connected individuals are on average substantially less likely to change opinion. We now ask whether analogous relationships emerge at the level of the entire interaction network.

Figure~\ref{fig:Pflip_netz} compares the weekly average flip probability with global network observables after standardization: the mean weighted degree $\langle k_w\rangle$, the weighted-degree heterogeneity $\mathrm{Var}(k_w)$, the average clustering coefficient $C$, and the mean weighted betweenness centrality $B_w$.

A clear monotonic relationship is observed for all four quantities. Weeks characterized by larger average connectivity or stronger degree heterogeneity exhibit significantly lower probabilities of opinion change, with Spearman correlations of $\rho_S=-0.64$ and $\rho_S=-0.77$, respectively. Conversely, the flip probability increases with both the mean clustering coefficient ($\rho_S=0.61$) and the average weighted betweenness centrality ($\rho_S=0.74$).

These correlations suggest that the global organization of the communication network modulates collective opinion stability. Highly heterogeneous networks, dominated by influential hubs, correspond to periods of increased opinion persistence, whereas more clustered or bridge-dominated network configurations are associated with greater opinion mobility. In particular, the strong dependence on weighted-degree heterogeneity indicates that not only the average number of interactions, but also the inequality in their distribution, plays a major role in determining how opinions evolve.

Together with the node-level analysis of Fig.~\ref{fig:flip_vs_degree}, these results reveal a multiscale relationship between network topology and opinion dynamics: connectivity reduces the probability of opinion change both locally, through the persistence of highly connected individuals, and globally, through changes in the overall architecture of the interaction network.

\section{Kinetic Ising models for opinion dynamics}\label{sec:Mod}

We consider a population of individuals whose opinion at time $t$ is represented by a binary spin variable
$s_i(t)\in\{-1,+1\}$.
The interaction network is described by a weighted adjacency matrix
$J_{ij}(t)$,
where the weight encodes the intensity of interactions between users during week $t$.
For each node, we define the normalized local social field

\begin{equation}
u_i(t)
=
\frac{\sum_j J_{ij}(t)s_j(t)}
{\sum_j J_{ij}(t)},
\label{eq:localfield}
\end{equation}

which measures the average opinion of the neighborhood {in the retweet network} weighted by interaction strength.
Using the normalized field allows the influence experienced by each individual to remain bounded between $-1$ and $1$, independently of its degree.

Opinion updates are modeled through a kinetic Ising heat-bath dynamics \cite{glauber1963}.
The probability that individual $i$ adopts opinion $+1$ at the next observation time is

\begin{equation}
P\!\left(s_i(t+1)=+1\right)
=
\sigma\!\left(H_i(t)\right)
=
\frac{1}{1+\exp[-H_i(t)]},
\label{eq:heatbath}
\end{equation}

where $\sigma(\cdot)$ denotes the logistic sigmoid function and
$H_i(t)$ is the effective social field acting on node $i$.
Equation~(\ref{eq:heatbath}) corresponds to the standard heat-bath transition probability of the kinetic Ising model \cite{glauber1963,newman1999monte}. 

The different microscopic models considered in this work correspond to different parameterizations of the effective field $H_i$.

\subsection{The kinetic Ising Models}

\paragraph{Model M1: social influence.}

The simplest model assumes that opinion changes are solely driven by the local social field,

\begin{equation}
H_i(t)
=
h
+
\beta\,u_i(t),
\label{eq:M1}
\end{equation}

where $h$ is an external field representing the intrinsic preference toward one of the two opinions, while $\beta$ measures the strength of social interactions. Larger values of $\beta$ correspond to a stronger tendency for individuals to align with the average opinion of their neighborhood.

\paragraph{Model M2: asymmetric response.}

To allow individuals holding opposite opinions to react differently to their neighborhood, independent effective fields are estimated for each current opinion state,

\begin{equation}
H_i(t)=
\begin{cases}
h_{+}+\beta_{+}\,u_i(t),
&
s_i(t)=+1,\\[2mm]
h_{-}+\beta_{-}\,u_i(t),
&
s_i(t)=-1.
\end{cases}
\label{eq:M2}
\end{equation}

This model tests whether the two opinion groups exhibit different intrinsic biases ($h_\pm$) or different susceptibilities to social influence ($\beta_\pm$).

\paragraph{Model M3: temporal persistence.}

Instead of fitting two independent response functions, persistence is introduced explicitly through a memory term,

\begin{equation}
H_i(t)
=
h
+
\beta\,u_i(t)
+
\lambda\,s_i(t),
\label{eq:M3}
\end{equation}

where $\lambda$ is a persistence (or temporal inertia) parameter. Positive values of $\lambda$ increase the probability that an individual retains its previous opinion independently of the current social field.

\paragraph{Model M4: degree-dependent persistence.}

Finally, we investigate whether persistence is an intrinsic property or instead emerges from network position. We therefore allow the persistence strength to increase with the weighted degree,

\begin{equation}
H_i(t)
=
h
+
\beta\,u_i(t)
+
\lambda_k\,s_i(t)\log\!\left(1+k_i(t)\right),
\label{eq:M4}
\end{equation}

where

\begin{equation}
k_i(t)=\sum_j J_{ij}(t)
\end{equation}

denotes the weighted degree of node $i$. The logarithmic dependence accounts for the broad degree distribution while preventing highly connected hubs from dominating the dynamics. In this formulation, persistence is no longer homogeneous across the population but increases with network connectivity: highly connected individuals
are expected to display stronger persistence against opinion changes.

Models M1--M4 therefore define a hierarchy of kinetic Ising models of increasing complexity. Starting from social influence alone (M1), they successively introduce asymmetric responses (M2), temporal persistence (M3), and degree-dependent persistence (M4), allowing us to identify the microscopic mechanisms underlying empirical opinion dynamics.

\subsection{Connection with continuous opinion dynamics}

In a previous article \cite{achitouv2026dmodddiffusionmodelopinion} a framework models opinion evolution as a continuous stochastic process governed by the Langevin equation was proposed (D-MODD):

\begin{equation}
dx
=
A(x)\,dt
+
\sqrt{2D(x)}\,dW_t,
\end{equation}

where $x$ denotes the continuous opinion of an individual, while $A(x)$ and $D(x)$ are the empirical drift and diffusion inferred directly from data. The kinetic Ising model considered here can be viewed as the binary counterpart of this continuous description.

Indeed, with the heat-bath update rule of Eq.\ref{eq:heatbath} implies that the conditional expectation of the spin satisfies

\begin{equation}
\langle s_i(t+1)\rangle
=
2P(+1)-1
=
\tanh\!\left(H_i(t)/2\right).
\end{equation}

Without requiring a weak-field approximation, the expected change of the
spin between two consecutive observations can be written exactly as
\begin{equation}
\Delta s_i
=
\langle s_i(t+1)\rangle
-
s_i(t)
=
-s_i(t)
+
\tanh\!\left(\frac{H_i(t)}{2}\right).
\end{equation}
Interpreting one discrete update as occurring over a time interval
$\Delta t$, Eq.~(11) can be embedded in continuous time as
\begin{equation}
\frac{d s_i}{dt}
=
\frac{1}{\Delta t}
\left[
-s_i
+
\tanh\!\left(\frac{H_i}{2}\right)
\right].
\end{equation}
Defining the corresponding characteristic rate
$\kappa \equiv 1/\Delta t$, we obtain
\begin{equation}
\frac{d s_i}{dt}
=
A(s_i)
=
\kappa
\left[
-s_i
+
\tanh\!\left(\frac{H_i}{2}\right)
\right].
\end{equation}
This expression has the form of a nonlinear drift equation and therefore
provides a direct connection with the drift term inferred in the D-MODD
framework. In particular, sufficiently close to $H_i=0$, one recovers the
linear expansion
\begin{equation}
A(s_i)
\simeq
-\kappa s_i
+
\frac{\kappa}{2}H_i
+
\mathcal{O}(H_i^3).
\end{equation}

The present work therefore provides the binary analogue of D-MODD: rather than reconstructing a continuous drift field, we infer the microscopic transition kernel governing binary opinion changes. Both approaches recover the underlying dynamics directly from empirical observations, differing only in the choice of state space.

\subsection{Sociological interpretations}
We first compare the predictive performance of the four kinetic Ising models. Table~\ref{tab:model_comparison} reports the average log-loss over all weekly fits together with the number of free parameters. Lower log-loss indicates better predictive accuracy.

\begin{table}[h]
\centering
\caption{Comparison of the four kinetic Ising models. The reported log-loss corresponds to the average predictive log-loss over all weekly fits.}
\label{tab:model_comparison}
\begin{tabular}{lcc}
\hline
Model & Parameters & Mean log-loss \\
\hline
M1: Social influence & 2 & 0.1103 \\
M2: Asymmetric response & 4 & 0.1048 \\
M3: Memory & 3 & 0.1056 \\
M4: Degree-dependent persistence & 3 & \textbf{0.1034} \\
\hline
\end{tabular}
\end{table}

The baseline model M1 already captures a large fraction of the observed opinion dynamics. Introducing temporal persistence  improves predictive performance, highlighting the importance of individual inertia. Allowing persistence to depend on node degree (M4) further reduces the prediction error and achieves the best performance while using only three parameters, outperforming the four-parameter asymmetric model M2. This suggests that the improvement obtained by allowing independent
response functions for the two opinion groups can largely be captured by
an explicit persistence mechanism. Allowing this persistence to depend on
node degree (M4) further improves the predictive performance, indicating
that network connectivity provides useful information for describing the
heterogeneity of opinion persistence.


\subsubsection{Dynamical evolution of the model parameters}

\begin{figure}
    \centering
    \includegraphics[width=1\linewidth]{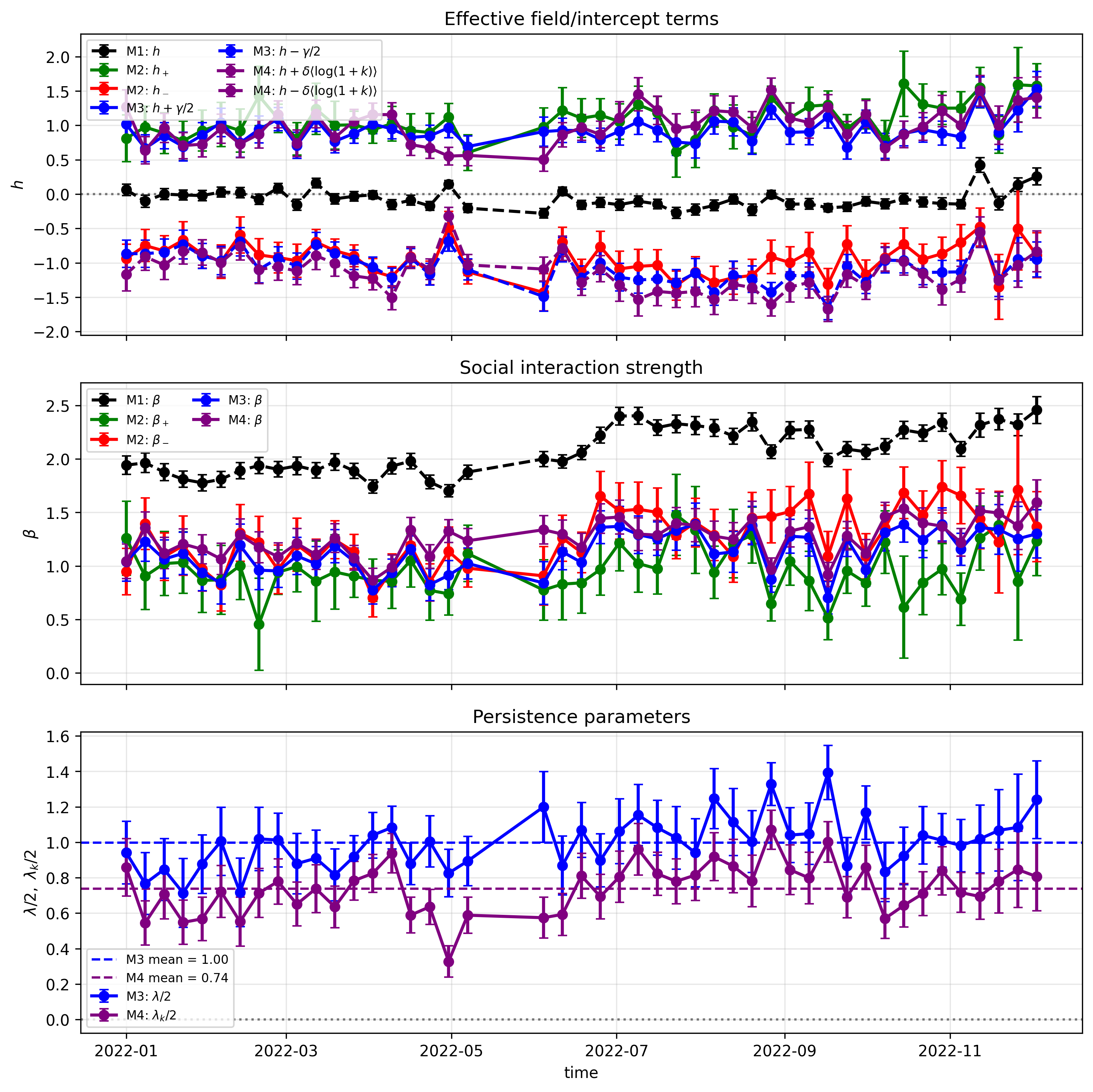}
    \caption{
Weekly evolution of the kinetic Ising parameters inferred from the four models. 
}
    \label{fig:model_parameters}
\end{figure}

Figure~\ref{fig:model_parameters} reports the weekly evolution of the effective field $h$, the interaction strength $\beta$, and the persistence parameters inferred from the four kinetic Ising models.

\paragraph{Effective field.}

The field-only model (M1) yields an external field that remains close to zero throughout the year. This result is expected because the model pools together users currently holding opposite opinions. When persistence is strong, the conditional effective fields satisfy approximately
\[
h_{+}\simeq -h_{-},
\]
so that averaging over both populations nearly cancels the intercept contribution, leading to an apparent field $h\simeq0$.

Conditioning on the current opinion state (M2) reveals this hidden persistence. The inferred effective fields are of similar magnitude but opposite sign,
\[
h_{+}\approx -h_{-}, \qquad |h_{\pm}|\approx1,
\]
showing that individuals have an intrinsic tendency to remain in their current opinion state, consistent with the well-documented persistence of political attitudes~\cite{sears1999evidence}.

Model M3 explains the same separation through a single persistence parameter rather than two independent intercepts,
\[
h_{+}=h+\frac{\lambda}{2},
\qquad
h_{-}=h-\frac{\lambda}{2},
\]
where $\lambda$ measures temporal persistence. The close overlap between the effective fields inferred by M2 and M3 demonstrates that the apparent asymmetry between the two opinion groups is almost entirely explained by memory. This provides a considerably more parsimonious description of the dynamics while maintaining essentially the same predictive performance.

Finally, Model M4 replaces the homogeneous persistence parameter by a degree-dependent persistence,
\[
h_{\pm}
=
h
\pm
\frac{\lambda_k}{2}\log(1+k_i),
\]
where $\lambda_k$ quantifies how persistence increases with node connectivity.
The inferred effective fields remain remarkably close to those of M2 and M3,
indicating that node degree captures a substantial fraction of the
heterogeneity associated with opinion persistence.

The inferred effective fields remain remarkably close to those of M2 and M3,
indicating that node degree captures a substantial fraction of the
heterogeneity associated with opinion persistence. In contrast to M3, where
the same persistence strength $\lambda$ is assigned to all users, M4
redistributes this contribution according to connectivity through the factor
$\log(1+k_i)$. Consequently, the smaller value of $\lambda_k$ observed in
Fig.~4 does not correspond to a weaker persistence, since its effective
contribution is amplified for increasingly connected users.

\paragraph{Social interaction strength.}

The coupling $\beta$ measures the response of individuals to the local social field $u_i(t)$. In all four models $\beta$ remains positive throughout the observation period, indicating that users preferentially align with the opinions expressed in their local neighborhood, consistently with the basic social-influence mechanism underlying classical models of opinion
dynamics~\cite{degroot_reaching_1974,holley_ergodic_1975,castellano2009}.

The field-only model (M1) consistently produces the largest coupling strengths. This systematic overestimation (compared to M2,M3 \& M4) arises because, in the absence of an explicit persistence mechanism, the social field partially absorbs the temporal inertia present in the data. Since users are typically surrounded by like-minded neighbors, the local field $u_i(t)$ is positively correlated with the current opinion $s_i(t)$. Consequently, the fitted coupling simultaneously captures genuine social influence and persistence, artificially inflating the estimate of $\beta$.

Introducing persistence explicitly (M2 and M3) substantially reduces the inferred interaction strength while leaving its temporal evolution largely unchanged. This indicates that social influence and temporal persistence correspond to distinct mechanisms that can be disentangled statistically. Interestingly, allowing persistence to depend on node degree (M4) produces
coupling strengths very similar to those inferred with M3. This indicates
that introducing degree dependence in the persistence term does not
substantially modify the inferred strength of social interactions.

\paragraph{Persistence parameters.}

The persistence parameters reported in the third panel reveal a stable
temporal inertia throughout the year. The homogeneous persistence inferred
by M3 is consistently larger than the degree-dependent coefficient inferred
by M4 because the latter is multiplied by $\log(1+k_i)$, allowing the
effective persistence contribution to vary with connectivity. Accordingly,
M4 assigns stronger persistence to highly connected users than to peripheral
users, consistently with the empirical decrease of the flip probability with
node degree observed in Fig.~\ref{fig:flip_vs_degree}. 

This result reveals a strong correlation between network connectivity and
opinion stability at the population level, while individual trajectories may
provide notable exceptions, such as initially poorly connected but highly
committed ``super-spreaders'' discussed above.


\subsubsection{Correlations with the network topology}

Table~\ref{tab:persistence_network_corr} summarizes the Pearson correlations between the inferred persistence magnitude and the principal global network observables. Remarkably, all four models exhibit the same qualitative behavior: persistence increases with both the mean weighted degree and the heterogeneity of the degree distribution, while it decreases with the mean clustering coefficient and the average weighted betweenness.

In the field-only model (M1), the magnitude of the effective field $|h|$ already displays a moderate positive correlation with weighted-degree heterogeneity ($r=0.50$) and a negative correlation with clustering ($r=-0.32$). Similar trends are recovered in the spin-conditioned model (M2), although the correlations are weaker and differ slightly between the two opinion states.

The same behavior persists in the explicit persistence models. The homogeneous persistence parameter $\lambda$ inferred in M3 increases with both the average weighted degree ($r=0.34$) and the degree heterogeneity ($r=0.42$), while decreasing with clustering ($r=-0.42$) and weighted betweenness ($r=-0.51$). Finally, the degree-dependent persistence inferred by M4,
\[
\frac{\lambda_k}{2}\left\langle\log(1+k_i)\right\rangle,
\]
exhibits the strongest correlation with the mean weighted degree ($r=0.64$) while maintaining the same dependence on the remaining network observables.

These results indicate that periods characterized by highly connected and
heterogeneous interaction networks are associated with stronger temporal
inertia, whereas more clustered or bridge-dominated networks exhibit weaker
persistence. Combined with the node-level analysis of
Fig.~\ref{fig:flip_vs_degree}, these results reveal a consistent correlation
between network connectivity and opinion stability across both individual
and global network scales. The stronger correlations obtained for M4 further
show that node connectivity provides a relevant observable for capturing
heterogeneity in opinion persistence. 
Importantly, these correlations do not establish the direction of causality:
network connectivity may influence opinion persistence, but persistent or
highly committed behavior may also shape the network connectivity of
individuals. Moreover, these correlations correspond to an average population-level
tendency, and departures may occur at the individual level; initially poorly
connected but highly committed ``super-spreaders'', as discussed above,
provide a possible exception.


Table~\ref{tab:beta_network_corr} reports the correlations between the inferred social interaction strength $\beta$ and the global network observables. In all models, stronger social coupling is associated with more heterogeneous and more connected interaction networks, while it decreases with both the mean clustering coefficient and the average weighted betweenness. This indicates that stronger inferred social alignment is associated with
more highly connected and heterogeneous networks than with more locally
clustered or bridge-dominated communication structures.

The strongest correlations are obtained for the field-only model (M1), where $\beta$ exhibits a strong positive correlation with both the weighted-degree heterogeneity ($r=0.73$) and the mean weighted degree ($r=0.65$), together with strong negative correlations with clustering ($r=-0.55$) and weighted betweenness ($r=-0.70$). However, these correlations decrease substantially once temporal persistence is introduced explicitly in Models M2--M4.

This behavior has a natural interpretation. In the absence of a persistence
term, Model M1 forces the social interaction strength to account
simultaneously for alignment with neighboring opinions and for the tendency
of individuals to remain in their current opinion state. Since users are
predominantly connected to like-minded neighbors, the local field $u_i(t)$
is itself correlated with temporal persistence, leading M1 to absorb part of
this persistence into the effective coupling strength and, consequently, to
stronger apparent correlations with the network topology.

By contrast, Models M3 and M4 explicitly separate social alignment from
temporal inertia. As a result, the correlations between $\beta$ and the
network topology become significantly weaker, while the persistence
parameters exhibit stronger correlations with several of the same network
observables (Table~\ref{tab:persistence_network_corr}). In particular, the
degree-dependent persistence inferred by M4 is much more strongly correlated
with the mean weighted degree than the corresponding interaction strength. These results indicate that, once temporal persistence is explicitly
accounted for, network topology is more strongly correlated with the
persistence component of the dynamics than with the inferred strength of
social alignment.

\begin{table}[t]
\centering
\caption{Pearson correlations between the inferred persistence magnitude and global
network observables.}
\label{tab:persistence_network_corr}
\begin{tabular}{lcccc}
\hline
Model &
$\mathrm{Var}(k_w)$ &
$\langle k_w\rangle$ &
$C$ &
$\langle B_w\rangle$\\
\hline
M1: $|h|$
& 0.497
& 0.225
& $-0.318$
& $-0.424$ \\

M2: $|h_+|$
& 0.380
& 0.153
& $-0.155$
& $-0.268$ \\

M2: $|h_-|$
& 0.092
& 0.238
& $-0.287$
& $-0.335$ \\

M3: $\lambda$
& 0.417
& 0.339
& $-0.419$
& $-0.508$ \\

M4: $\lambda_k\langle\log(1+k)\rangle$
& \textbf{0.459}
& \textbf{0.641}
& $-\mathbf{0.390}$
& $-\mathbf{0.504}$ \\
\hline
\end{tabular}
\end{table}

\begin{table}[t]
\centering
\caption{Pearson correlations between the inferred social interaction strength $\beta$ and global network observables. Positive values indicate stronger social alignment in more connected or heterogeneous networks, whereas negative values indicate that coupling decreases with increasing clustering or weighted betweenness.}
\label{tab:beta_network_corr}
\begin{tabular}{lcccc}
\hline
Model &
$\mathrm{Var}(k_w)$ &
$\langle k_w\rangle$ &
$C$ &
$\langle B_w\rangle$\\
\hline
M1: $\beta$
& \textbf{0.732}
& \textbf{0.654}
& $-\mathbf{0.549}$
& $-\mathbf{0.700}$ \\

M2: $\beta_{+}$
& 0.141
& 0.215
& $-0.075$
& $-0.074$ \\

M2: $\beta_{-}$
& 0.641
& 0.540
& $-0.439$
& $-0.542$ \\

M3: $\beta$
& 0.505
& 0.474
& $-0.296$
& $-0.400$ \\

M4: $\beta$
& 0.440
& 0.208
& $-0.280$
& $-0.358$ \\
\hline
\end{tabular}
\end{table}

\section{Testing the Ising dynamics}\label{sec:MCMC}

\subsection{Generative kinetic Ising simulations}

To assess whether the inferred transition kernels reproduce the collective dynamics beyond one-step prediction, we performed autonomous simulations using a non-stationary kinetic Ising model with parallel heat-bath updates. The empirical retweet network at week $t$ defines the interaction matrix $J(t)$, while the transition kernel is parameterized by the coefficients inferred from the weekly logistic regressions (Models M1--M4). The simulation is initialized from the empirical spin configuration at the first observation week, after which the dynamics evolve autonomously.

At each week $t$, the normalized local field acting on node $i$ is computed from the current simulated configuration,

\[
u_i(t)=
\frac{\sum_j J_{ij}(t)s_j(t)}
{\sum_j J_{ij}(t)},
\]

where the sum runs only over active nodes present in the empirical graph at week $t$. Unlike the one-step validation, the local field is therefore entirely determined by the simulated spins and is no longer taken from the empirical data.

For each node, an effective Ising field $H_i(t)$ is constructed according to the considered model,

\[
\begin{aligned}
\text{M1: }&
H_i
=
h+\beta\,u_i,\\[1mm]
\text{M2: }&
H_i=
\begin{cases}
h_{+}+\beta_{+}u_i,
&
s_i(t)=+1,\\[2mm]
h_{-}+\beta_{-}u_i,
&
s_i(t)=-1,
\end{cases}\\[3mm]
\text{M3: }&
H_i
=
h+\beta\,u_i+\lambda\,s_i,\\[1mm]
\text{M4: }&
H_i
=
h+\beta\,u_i
+
\lambda_k\,s_i\log\!\left(1+k_i\right),
\end{aligned}
\]

where $k_i$ denotes the weighted degree of node $i$.

Spins are then updated simultaneously according to the heat-bath (Glauber) transition probability

\[
P\!\left(s_i(t+1)=+1\right)
=
\frac{1}
{1+\exp\!\left[-2H_i(t)\right]},
\]

with

\[
P\!\left(s_i(t+1)=-1\right)
=
1-
P\!\left(s_i(t+1)=+1\right).
\]

The interaction matrix $J(t)$ and the corresponding model parameters are updated every week using the empirical interaction network and the coefficients inferred from the data. The resulting dynamics therefore constitute a non-stationary kinetic Ising process whose interactions and effective fields evolve according to the observed temporal evolution of the social network.

\subsection{Results}

\begin{figure}
    \centering
    \includegraphics[width=1\linewidth]{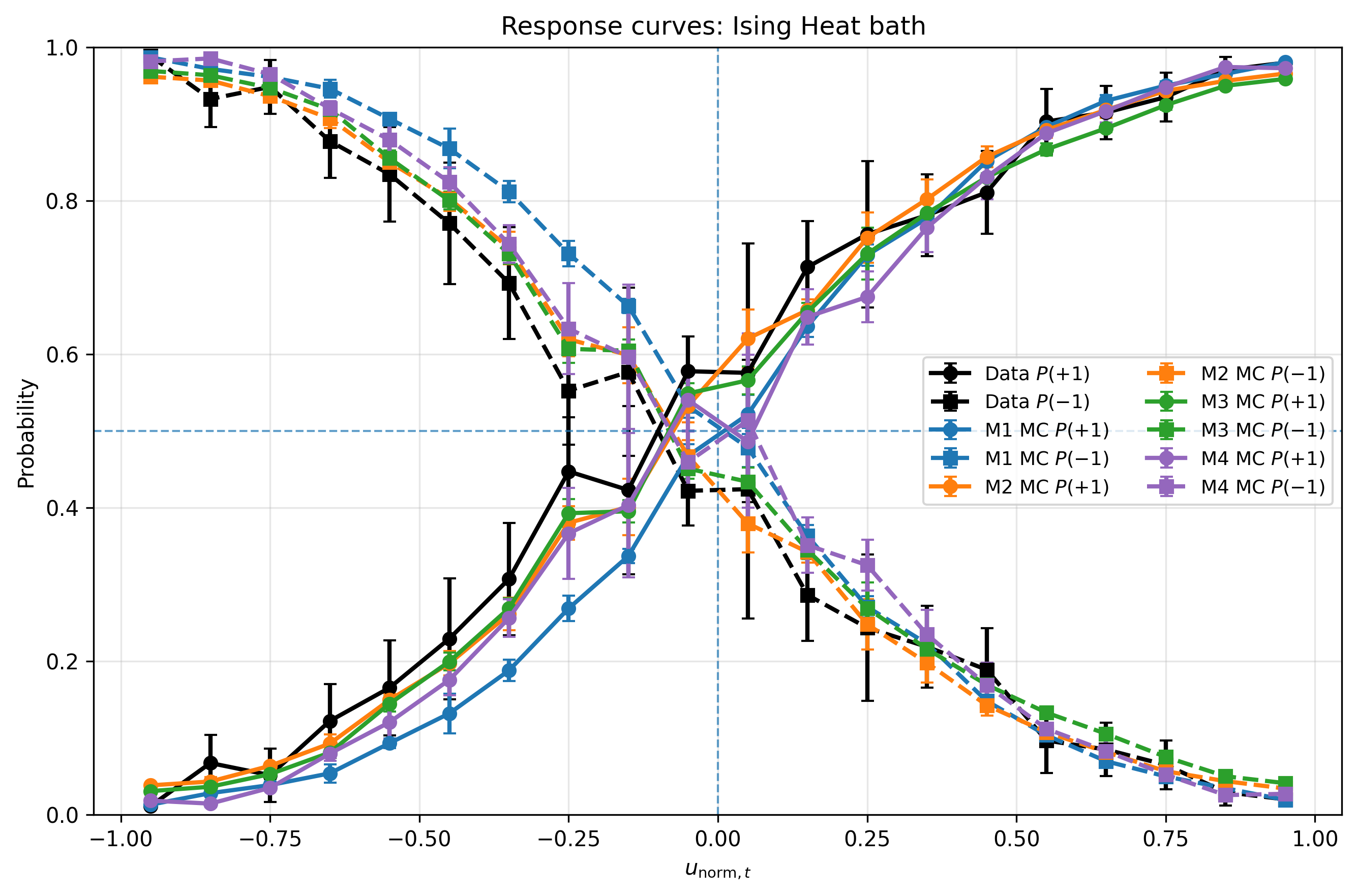}
    \caption{Empirical opinion-update kernel and kinetic Ising predictions. Black symbols show the empirical conditional probabilities ($P(s_i(t+1)=\pm1,|,u_{\mathrm{norm},t}$)) computed from the Twitter data as a function of the normalized local field ($u_{\mathrm{norm},t}$). Colored curves correspond to Monte Carlo simulations of the four inferred kinetic Ising models (M1-M4) using the weekly estimated parameters.}
    \label{fig:response_curves}
\end{figure}

We now investigate whether the microscopic dynamics of opinion changes can be described by a kinetic Ising model. We first compare the empirical transition probabilities with the heat-bath response functions predicted by the four models introduced in Sec.~\ref{sec:Mod}. 

Figure~\ref{fig:response_curves} compares the empirical probability of adopting each opinion with the heat-bath response curves obtained from the four inferred kinetic Ising models. The black curves correspond to the empirical conditional probabilities,
\[
P\!\left(s_i(t+1)=\pm1 \mid u_i(t)\right),
\]
while the colored curves are obtained from Monte Carlo simulations using the inferred weekly parameters.

All four models reproduce the characteristic sigmoidal dependence of the transition probabilities on the normalized local field \(u_{\mathrm{i},t}\), demonstrating that binary opinion updates are well described by the heat-bath dynamics of a kinetic Ising model. As expected, the probability of adopting opinion \(+1\) increases monotonically with the local field, while the probability of adopting opinion \(-1\) decreases symmetrically.

The differences between models are most pronounced around the transition region \(u_{\mathrm{norm}}\approx0\), where opinion changes are most likely. The field-only model (M1) systematically underestimates the persistence of opinions, producing a smoother transition than observed empirically. Introducing state-dependent effective fields (M2) improves the agreement, while the explicit persistence model (M3) achieves a comparable fit using a single memory parameter. Finally, the degree-dependent persistence model (M4) provides the closest overall agreement with the empirical response curves, particularly around the critical transition region where the competition between social influence and individual persistence is strongest.

\begin{figure}
    \centering
    \includegraphics[width=1\linewidth]{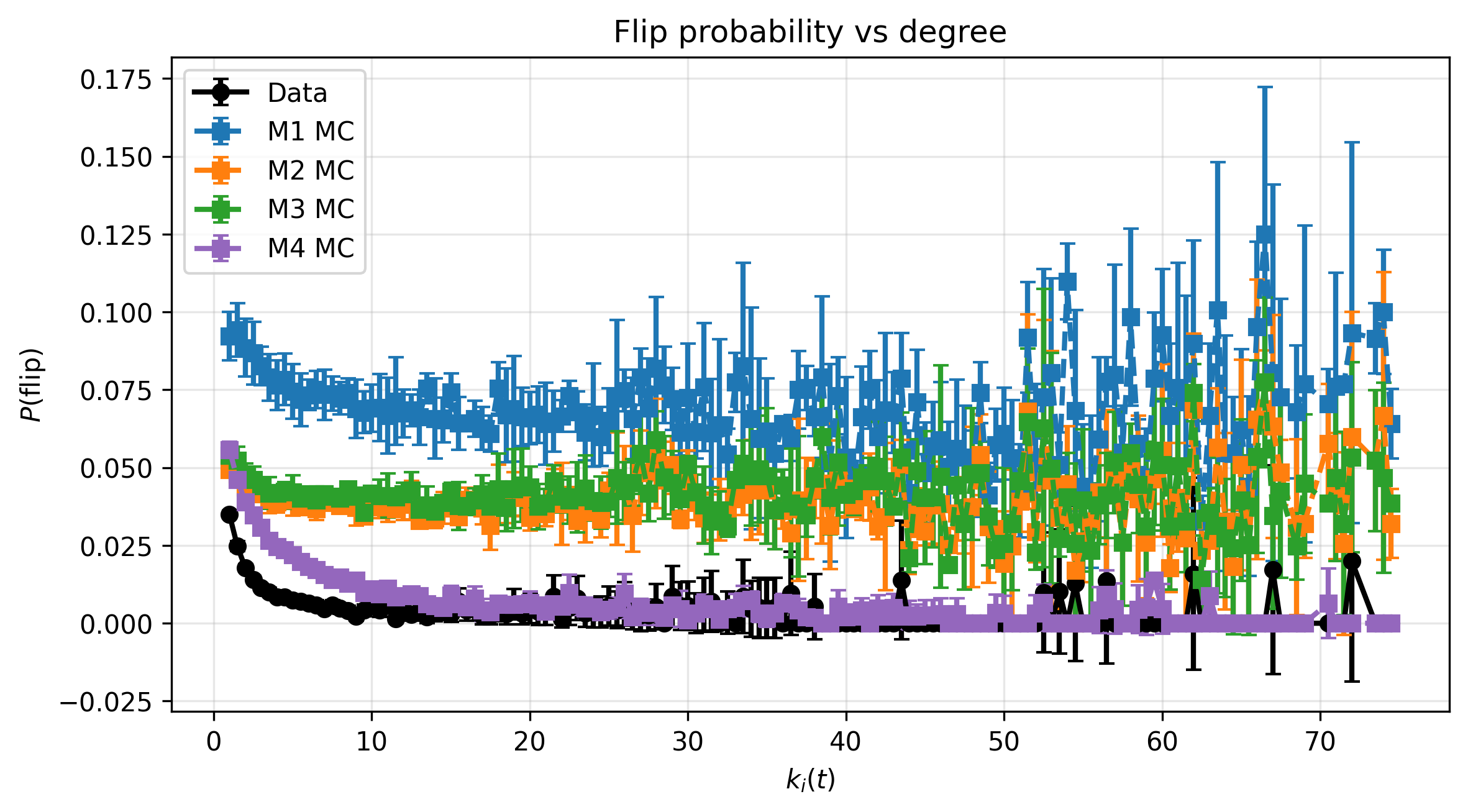}
    \caption{Opinion flip probability as a function of weighted degree. Black symbols show the empirical probability that a user changes opinion between two consecutive weeks, averaged over all weeks, as a function of its weighted degree ($k_i(t)$). Colored curves correspond to Monte Carlo simulations of the four kinetic Ising models. }
    \label{fig:flip_vs_degree}
\end{figure}

Another central question is whether persistence is an intrinsic property of individuals or whether it emerges from their position in the interaction network. To address this question, Fig.~\ref{fig:flip_vs_degree} reports the empirical probability that a user changes opinion between two consecutive weeks as a function of its weighted degree.

The empirical data reveal a clear monotonic decrease of the flip probability with increasing degree. Low-degree users change opinion relatively frequently, with flip probabilities of approximately (3.5\%-1\% for k in [1,5]), whereas highly connected users almost never switch opinion. This observation suggests that influential users occupy considerably more stable opinion states than peripheral users.

The four kinetic Ising models produce markedly different predictions. Because M1 does not include any persistence mechanism, it systematically overestimates the probability of opinion changes over the entire degree range. Introducing spin-dependent intercepts (M2) or an explicit memory parameter (M3) substantially reduces the flip probability, but both models remain largely independent of node degree and therefore fail to reproduce the strong empirical decrease observed for highly connected users.

In contrast, the degree-dependent persistence model (M4) captures both the magnitude and the functional form of the empirical relationship. By making the persistence strength proportional to ($\log(1+k_i)$ ), M4 naturally predicts that highly connected individuals become increasingly resistant to opinion changes while low-degree users remain comparatively volatile.
Note that the remaining discrepancy at low degree may originate from the specific
logarithmic dependence on connectivity assumed in M4. The function
$\log(1+k_i)$ may impose too strong a degree dependence in this regime,
suggesting that a more strongly compressive or saturating functional form
could further improve the agreement with the empirical dynamics.

Overall, these results provide direct evidence that temporal persistence is not
a homogeneous behavioral property but is strongly associated with the structure
of the interaction network. In particular, the empirical dependence of the
flip probability on node degree motivates the introduction of a
degree-dependent persistence term in the kinetic Ising dynamics for
heterogeneous networks. Model M4 provides the closest reproduction of the
empirical dynamics, substantially improving the agreement with the observed
flip probability compared with the other microscopic update rules. This
effective degree dependence should not, however, be interpreted as a causal
effect of connectivity on persistence, as both individual characteristics and
population composition may contribute to the observed relationship.

\section{Conclusion}\label{sec:conclu}

We introduced a data-driven framework to infer the microscopic mechanisms
governing online opinion dynamics using a kinetic Ising model. Rather than
assuming interaction parameters, we reconstruct the transition kernel
directly from observed opinion updates in a year-long social media dataset.
This allows each parameter of the model to be interpreted and validated
empirically.

Our results reveal three principal mechanisms underlying opinion evolution.
First, opinion updates exhibit the sigmoid response to the local social field
predicted by kinetic Ising dynamics, providing quantitative evidence for a
systematic alignment between individual opinion updates and the opinions of
their local neighborhood. Second, explicitly accounting for temporal persistence captures much of the additional structure obtained by allowing independent response functions for the two opposing opinion groups, but with fewer degrees of freedom in the model parameterization, indicating that opinion inertia constitutes an essential component of the microscopic dynamics. Finally, persistence exhibits a strong correlation with the
structure of the interaction network: periods characterized by highly
connected and heterogeneous networks, as well as highly connected users,
are associated with greater opinion inertia, whereas more clustered networks
are associated with weaker persistence. Importantly, these correlations do
not establish the direction of causality: network connectivity may influence
opinion persistence, while persistent or highly committed behavior may
conversely shape the connectivity of individuals.

Beyond parameter estimation, the inferred kinetic Ising models reproduce the
empirical microscopic response curves and generate realistic macroscopic
opinion dynamics. In particular, the degree-dependent persistence model
provides the closest reproduction of the empirical dynamics, supporting
connectivity-dependent persistence as an effective description of the
observed heterogeneity in opinion stability. These results demonstrate that
a small number of interpretable mechanisms are sufficient to capture a
substantial fraction of the observed dynamics.

Finally, our approach naturally complements the continuous D-MODD framework
\cite{achitouv2026dmodddiffusionmodelopinion}. Whereas D-MODD reconstructs
the drift and diffusion of continuous opinion trajectories, the present work
identifies the binary microscopic transition kernel governing discrete
opinion changes. Together, these two frameworks provide a unified empirical
methodology for inferring stochastic opinion dynamics across different
representations of opinion states.

More broadly, this work establishes a bridge between statistical physics and
computational social science. While previous applications of Ising models to
opinion dynamics typically relied on assumed interaction rules, our framework
infers these rules directly from empirical data and relates the inferred
parameters to measurable network properties. This opens the possibility of
studying how changes in platform design, recommendation algorithms,
moderation policies, or network interventions are reflected in the
microscopic transition kernel and, consequently, in the collective evolution
of public opinion.

\noindent\textbf{Data availability.}
The anonymized opinion trajectories and interaction graphs used in this study will be made publicly available at \url{https://github.com/IxandraAchitouv/Kinetic_Ising_opinion_dyn.git}, upon publication along with the codes used to generate the figures.

\bibliography{reference}

@article{glauber1963,
  author = {Roy J. Glauber},
  title = {Time-Dependent Statistics of the Ising Model},
  journal = {Journal of Mathematical Physics},
  volume = {4},
  pages = {294--307},
  year = {1963}
}

@book{newman1999monte,
  title     = {Monte Carlo Methods in Statistical Physics},
  author    = {Newman, M. E. J. and Barkema, G. T.},
  year      = {1999},
  publisher = {Oxford University Press},
  address   = {Oxford},
  isbn      = {9780198517962},
  doi       = {10.1093/oso/9780198517962.001.0001}
}

@misc{peralta2022opiniondynamicssocialnetworks,
  title={Opinion dynamics in social networks: From models to data},
  author={Antonio F. Peralta and J{\'a}nos Kert{\'e}sz and Gerardo I{\~n}iguez},
  year={2022},
  eprint={2201.01322},
  archivePrefix={arXiv},
  primaryClass={physics.soc-ph},
  url={https://arxiv.org/abs/2201.01322}
}

@article{gaumont_reconstruction_2018,
	title = {Reconstruction of the socio-semantic dynamics of political activist {Twitter} networks—{Method} and application to the 2017 {French} presidential election},
	volume = {13},
	copyright = {All rights reserved},
	issn = {1932-6203},
	url = {http://dx.plos.org/10.1371/journal.pone.0201879},
	doi = {10.1371/journal.pone.0201879},
	language = {en},
	number = {9},
	urldate = {2018-09-19},
	journal = {PLOS ONE},
	author = {Gaumont, Noé and Panahi, Maziyar and Chavalarias, David},
	editor = {Araujo, Nuno},
	month = sep,
	year = {2018},
	pages = {e0201879},
}

@article{amblard_role_2004,
	title={The role of network topology on extremism propagation with the relative agreement opinion dynamics},
   volume={343},
   ISSN={0378-4371},
   url={http://dx.doi.org/10.1016/j.physa.2004.06.102},
   DOI={10.1016/j.physa.2004.06.102},
   journal={Physica A: Statistical Mechanics and its Applications},
   publisher={Elsevier BV},
   author={Amblard, Frédéric and Deffuant, Guillaume},
   year={2004},
   month=Nov, pages={725–738} }

@article{perrier_phase_2024,
	title = {Phase coexistence in the fully heterogeneous {Hegselmann}–{Krause} opinion dynamics model},
	volume = {14},
	issn = {2045-2322},
	url = {https://www.nature.com/articles/s41598-023-50463-z},
	doi = {10.1038/s41598-023-50463-z},
	language = {en},
	number = {1},
	urldate = {2024-02-12},
	journal = {Scientific Reports},
	author = {Perrier, Rémi and Schawe, Hendrik and Hernández, Laura},
	month = jan,
	year = {2024},
	pages = {241},
}

@article{galam_contrarian_2004,
	title={Contrarian deterministic effects on opinion dynamics: “the hung elections scenario”},
   volume={333},
   ISSN={0378-4371},
   url={http://dx.doi.org/10.1016/j.physa.2003.10.041},
   DOI={10.1016/j.physa.2003.10.041},
   journal={Physica A: Statistical Mechanics and its Applications},
   publisher={Elsevier BV},
   author={Galam, Serge},
   year={2004},
   month=Feb, pages={453–460} }

@article{degroot_reaching_1974,
	title = {Reaching a {Consensus}},
	volume = {69},
	issn = {0162-1459, 1537-274X},
	url = {http://www.tandfonline.com/doi/abs/10.1080/01621459.1974.10480137},
	doi = {10.1080/01621459.1974.10480137},
	language = {en},
	number = {345},
	urldate = {2025-11-27},
	journal = {Journal of the American Statistical Association},
	author = {Degroot, Morris H.},
	month = mar,
	year = {1974},
	pages = {118--121},
}

@article{friedkin_social_1990,
	title = {Social influence and opinions},
	volume = {15},
	issn = {0022-250X, 1545-5874},
	url = {http://www.tandfonline.com/doi/abs/10.1080/0022250X.1990.9990069},
	doi = {10.1080/0022250X.1990.9990069},
	language = {en},
	number = {3-4},
	urldate = {2025-11-27},
	journal = {The Journal of Mathematical Sociology},
	author = {Friedkin, Noah E. and Johnsen, Eugene C.},
	month = jan,
	year = {1990},
	pages = {193--206},
}

@article{deffuant_mixing_2000,
	title = {Mixing beliefs among interacting agents},
	volume = {03},
	issn = {0219-5259, 1793-6802},
	url = {https://www.worldscientific.com/doi/abs/10.1142/S0219525900000078},
	doi = {10.1142/S0219525900000078},
	language = {en},
	number = {01n04},
	urldate = {2025-11-27},
	journal = {Advances in Complex Systems},
	author = {Deffuant, Guillaume and Neau, David and Amblard, Frederic and Weisbuch, Gérard},
	month = jan,
	year = {2000},
	pages = {87--98},
}

@article{hegselmann2002,
  title   = {Opinion dynamics and bounded confidence models, analysis, and simulation},
  author  = {Hegselmann, Rainer and Krause, Ulrich},
  journal = {Journal of Artificial Societies and Social Simulation},
  volume  = {5},
  number  = {3},
  pages   = {1--33},
  year    = {2002},
  url     = {http://jasss.soc.surrey.ac.uk/5/3/2.html}
}

@article{holley_ergodic_1975,
	title = {Ergodic {Theorems} for {Weakly} {Interacting} {Infinite} {Systems} and the {Voter} {Model}},
	volume = {3},
	issn = {0091-1798},
	url = {https://projecteuclid.org/journals/annals-of-probability/volume-3/issue-4/Ergodic-Theorems-for-Weakly-Interacting-Infinite-Systems-and-the-Voter/10.1214/aop/1176996306.full},
	doi = {10.1214/aop/1176996306},
	number = {4},
	urldate = {2025-11-27},
	journal = {The Annals of Probability},
	author = {Holley, Richard A. and Liggett, Thomas M.},
	month = aug,
	year = {1975},
}

@article{ising_beitrag_1925,
	title = {Beitrag zur {Theorie} des {Ferromagnetismus}},
	volume = {31},
	copyright = {http://www.springer.com/tdm},
	issn = {0044-3328},
	url = {http://link.springer.com/10.1007/BF02980577},
	doi = {10.1007/BF02980577},
	number = {1},
	urldate = {2025-11-27},
	journal = {Zeitschrift für Physik},
	author = {Ising, Ernst},
	month = feb,
	year = {1925},
	pages = {253--258},
}

@article{chavalarias_hashtags_2025,
	title = {From hashtags to hostility: global dynamics of climate denialism on {Twitter} in the post-{COVID} era},
	volume = {357},
	issn = {1778-7025},
	shorttitle = {From hashtags to hostility},
	url = {https://comptes-rendus.academie-sciences.fr/geoscience/articles/10.5802/crgeos.304/},
	doi = {10.5802/crgeos.304},
	number = {G1},
	urldate = {2025-09-10},
	journal = {Comptes Rendus. Géoscience},
	author = {Chavalarias, David and Bouchaud, Paul and Chomel, Victor and Panahi, Maziyar},
	year = {2025},
	pages = {369--387},
}

@article{Sasahara_2020,
   title={Social influence and unfollowing accelerate the emergence of echo chambers},
   volume={4},
   ISSN={2432-2725},
   url={http://dx.doi.org/10.1007/s42001-020-00084-7},
   DOI={10.1007/s42001-020-00084-7},
   number={1},
   journal={Journal of Computational Social Science},
   publisher={Springer Science and Business Media LLC},
   author={Sasahara, Kazutoshi and Chen, Wen and Peng, Hao and Ciampaglia, Giovanni Luca and Flammini, Alessandro and Menczer, Filippo},
   year={2020},
   month=sep, pages={381–402} }

@article{Liu_2023,
   title={Emergence of Polarization in Coevolving Networks},
   volume={130},
   ISSN={1079-7114},
   url={http://dx.doi.org/10.1103/PhysRevLett.130.037401},
   DOI={10.1103/physrevlett.130.037401},
   number={3},
   journal={Physical Review Letters},
   publisher={American Physical Society (APS)},
   author={Liu, Jiazhen and Huang, Shengda and Aden, Nathaniel M. and Johnson, Neil F. and Song, Chaoming},
   year={2023},
   month=jan }

@unpublished{climatoscope,
  TITLE = {{The new fronts of denialism and climate skepticism}},
  AUTHOR = {Chavalarias, David and Bouchaud, Paul and Chomel, Victor and Panahi, Maziyar},
  URL = {https://hal.science/hal-04103183},
  NOTE = {working paper or preprint},
  YEAR = {2023},
  MONTH = May,
  HAL_ID = {hal-04103183},
  HAL_VERSION = {v2},
}

@inproceedings{grover2016node2vec,
  title     = {node2vec: Scalable Feature Learning for Networks},
  author    = {Grover, Aditya and Leskovec, Jure},
  booktitle = {Proceedings of the 22nd ACM SIGKDD International Conference on Knowledge Discovery and Data Mining},
  pages     = {855--864},
  year      = {2016},
  publisher = {ACM},
  doi       = {10.1145/2939672.2939754},
  url       = {https://doi.org/10.1145/2939672.2939754}
}

@article{mcinnes2018umap,
  title   = {UMAP: Uniform Manifold Approximation and Projection for Dimension Reduction},
  author  = {McInnes, Leland and Healy, John and Melville, James},
  journal = {arXiv preprint arXiv:1802.03426},
  year    = {2018},
  url     = {https://arxiv.org/abs/1802.03426}
}

@article{castellano2009,
   title={Statistical physics of social dynamics},
   volume={81},
   ISSN={1539-0756},
   url={http://dx.doi.org/10.1103/RevModPhys.81.591},
   DOI={10.1103/revmodphys.81.591},
   number={2},
   journal={Reviews of Modern Physics},
   publisher={American Physical Society (APS)},
   author={Castellano, Claudio and Fortunato, Santo and Loreto, Vittorio},
   year={2009},
   month=May, pages={591–646} }

@misc{achitouv2026dmodddiffusionmodelopinion,
      title={D-MODD: A Diffusion Model of Opinion Dynamics Derived from Online Data}, 
      author={Ixandra Achitouv and David Chavalarias and Raphael Fournier-S'niehotta},
      year={2026},
      eprint={2601.16226},
      archivePrefix={Physica A in press, arXiv},
      primaryClass={physics.soc-ph},
      url={https://arxiv.org/abs/2601.16226}, 
}

@article{sears1999evidence,
  title   = {Evidence of the Long-Term Persistence of Adults' Political Predispositions},
  author  = {Sears, David O. and Funk, Carolyn L.},
  journal = {The Journal of Politics},
  volume  = {61},
  number  = {1},
  year    = {1999},
  doi     = {10.2307/2647773}
}

\end{document}